\documentclass[slac_one]{revtex4}
\usepackage{graphicx}
\usepackage{fancyhdr}
\usepackage{axodraw}
\usepackage{color}
\def\red#1{{\color{red}#1}}
\def\green#1{{\color{green}#1}}
\def\blue#1{{\color{blue}#1}}

\newcommand{\met}{\rlap{\,/}E_T}

\begin{document}

\title{  
{\small 2005 International Linear Collider Workshop - Stanford U.S.A.}\\ 
\vspace{12pt}
Contrasting Supersymmetry and Universal Extra Dimensions at Colliders }

\author{Marco Battaglia}
\affiliation{Lawrence Berkeley National Laboratory, Berkeley, CA 94720, USA}
\author{Asesh K.~Datta}
\affiliation{MCTP, University of Michigan, Ann Arbor, MI 48109-1040, USA}
\author{Albert De Roeck}
\affiliation{CERN, Geneva, Switzerland}
\author{Kyoungchul Kong\footnote{This talk was given by K.~Kong, 
describing past and ongoing work performed in collaboration with 
the other authors.}, Konstantin T.~Matchev}
\affiliation{Physics Department, University of Florida, Gainesville, FL 32611, USA}

\begin{abstract}
We contrast the experimental signatures of low energy supersymmetry 
and the model of Universal Extra Dimensions and discuss various
methods for their discrimination at hadron and lepton colliders.
We study the discovery reach of hadron colliders for level 2 
Kaluza-Klein modes, which would indicate the presence of extra dimensions.
We also investigate the possibility to differentiate the spins 
of the superpartners and KK modes by means of the asymmetry method of Barr.
We then review the methods for discriminating between the 
two scenarios at a high energy linear collider such as CLIC. 
We consider the processes of Kaluza-Klein muon pair production 
in universal extra dimensions in parallel to smuon pair 
production in supersymmetry.
We find that the angular distributions of the final state muons,
the energy spectrum of the radiative return photon and the 
total cross-section measurement are powerful discriminators
between the two models. 
\end{abstract}

\maketitle

\thispagestyle{fancy}

\section{INTRODUCTION} 

Supersymmetry (SUSY) and Extra Dimensions offer two different paths to a theory of new 
physics beyond the Standard Model (SM). They both address the hierarchy problem, 
play a role in a more fundamental theory aimed at unifying the SM with gravity, and 
offer a candidate particle for dark matter, compatible with present cosmology data.
If either supersymmetry or extra dimensions exist at the TeV scale, signals of new 
physics should be found by the ATLAS and CMS experiments at the Large Hadron 
Collider (LHC) at CERN. However, as we discuss below in Section~\ref{sec:hadron}, 
the proper interpretation of such discoveries may not
be straightforward at the LHC and may require the complementary data from an 
$e^+e^-$ collider such as CLIC~\cite{clic}.

A particularly interesting scenario of TeV-size extra dimensions is 
offered by the so called Universal Extra Dimensions (UED) model,
originally proposed in~\cite{Appelquist:2000nn}, where all SM
particles are allowed to freely propagate into the bulk. 
The case of UED bears interesting analogies to supersymmetry 
and sometimes has been referred to as  ``bosonic supersymmetry'' 
\cite{Cheng:2002ab}. In principle, disentangling UED and 
supersymmetry appears highly non-trivial at hadron 
colliders~\cite{Cheng:2002ab,UEDhadron}.
For each SM particle, both models predict the existence of
a partner (or partners) with identical interactions.
Unfortunately, the masses of these new particles are model-dependent
and cannot be used to unambiguously discriminate between the two 
theories\footnote{Notice that 
the recently proposed little Higgs models with $T$-parity 
\cite{Tparity} are reminiscent of UED, and they may also be 
confused with supersymmetry.}. Both theories have a good
dark matter candidate~\cite{UEDdm} and the typical collider 
signatures contain missing energy. One would therefore
like to have experimental discriminators which relies 
on the fundamental distinctions between the two models.
In what follows we shall discuss methods for experimental
discrimination between SUSY and UED. In Section~\ref{sec:hadron}
we discuss the case of hadron colliders~\cite{KMTev, KongTev, KMAPS, KongPheno},
while in Section~\ref{sec:clic} we consider a future
high energy $e^+e^-$ collider~\cite{Battaglia:2005zf}.

\section{SUSY-UED DISCRIMINATION AT HADRON COLLIDERS}
\label{sec:hadron}

\subsection{Discovery of the KK tower}
\label{sec:Z2}

One of the characteristic features of UED is the presence
of a whole tower of Kaluza-Klein (KK) partners, labelled by their KK level $n$.
In contrast, $N=1$ supersymmetry predicts a single superpartner for each SM particle.
One might therefore hope to discover the higher KK modes of UED
and thus prove the existence of extra dimensions. However, there 
are two significant challenges along this route. First, 
the masses of the higher KK modes are (roughly)
integer multiples of the masses of the $n=1$ KK partners, and as a result 
their production cross-sections are kinematically suppressed. Second, the
$n=2$ KK particles predominantly decay to $n=1$ KK modes, 
which amounts to a small correction to
the inclusive production of $n=1$ KK particles. Just as in the case of SUSY, 
because of the unknown momentum carried away by the dark matter candidate
at the end of the decay chain, one is unable to reconstruct individual
KK resonances. The only exceptions are the level $2$ KK gauge bosons, 
which may appear as high mass dijet or dilepton resonances, when they 
decay directly to SM fermions through loop suppressed couplings~\cite{radcor}.

\begin{figure*}[t]
\centering
\includegraphics[width=82mm]{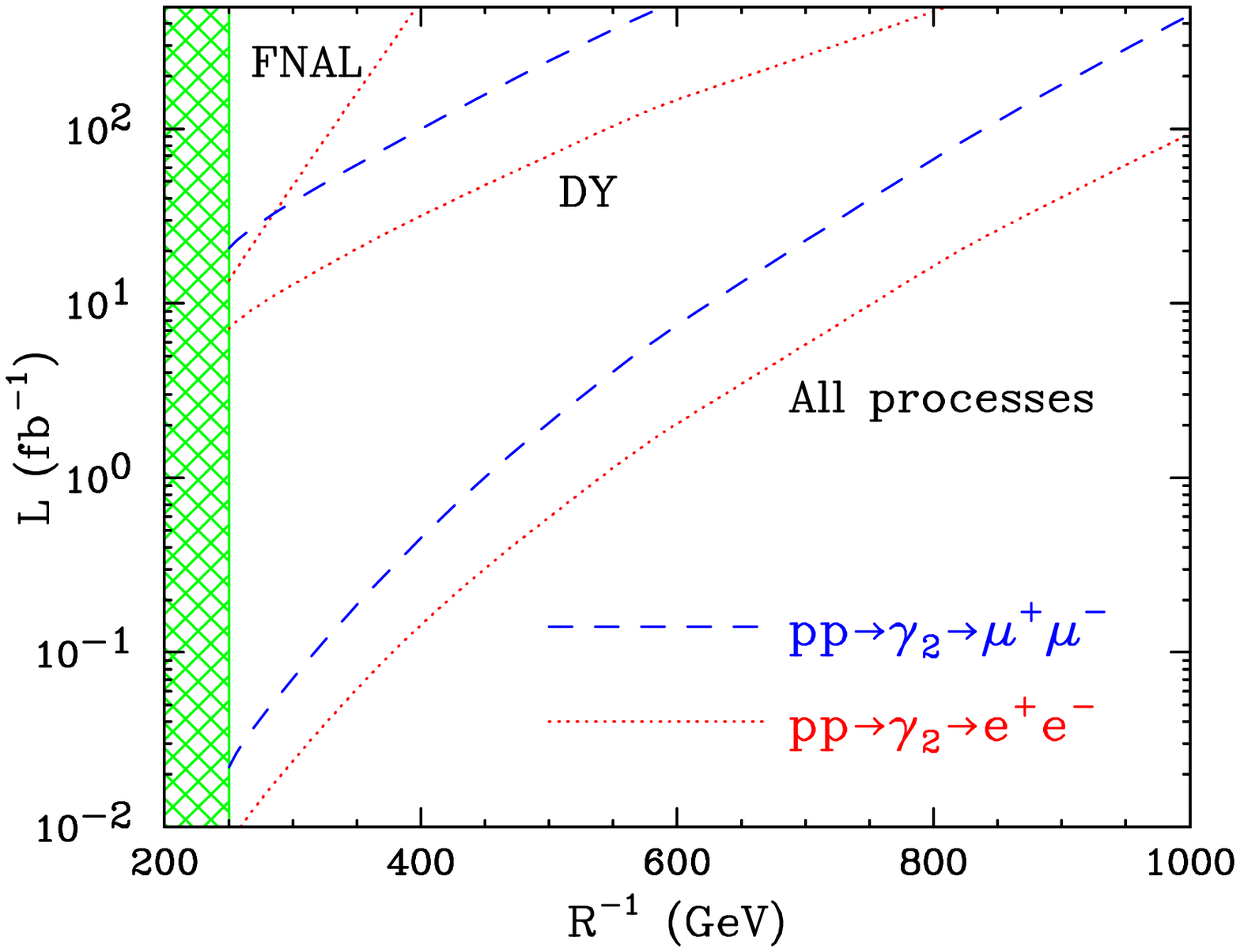}
\includegraphics[width=80mm]{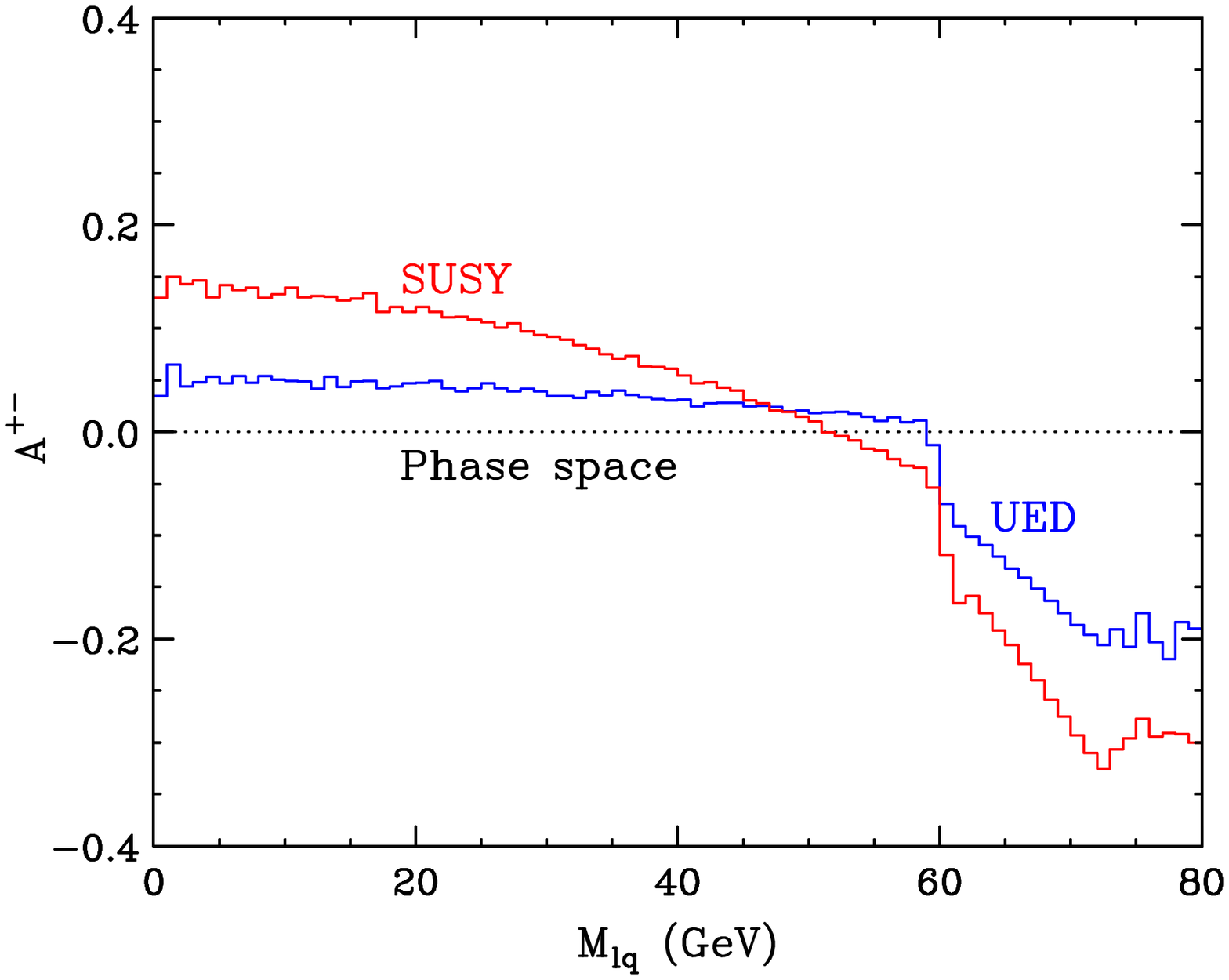}
\caption{Left: $5\sigma$ discovery reach of hadron colliders for the
level 2 Kaluza-Klein mode $\gamma_2$ of the photon.
The dashed (dotted) lines are for dimuon (dielectron) final states.
The lines labelled ``DY'' only include direct $\gamma_2$ production in Drell-Yan
processes, while the lower set of lines (labelled ``All processes'')
includes indirect $\gamma_2$ production from the decays of other level 2 KK modes.
For the Tevatron (FNAL) we only show the dielectron reach, including 
both direct and indirect production. Right: The charge asymmetry
defined in Eq.~(\ref{asymmetry}) in UED (blue) for $R^{-1}=500$ GeV,
and in SUSY (red).
The SUSY parameters have been chosen to get a matching spectrum.
The horizontal dotted line corresponds to the case when all spin 
correlations are neglected and particles decay according to pure 
phase space.}
\label{fig:reach_B2}
\end{figure*}

We studied the prospects for observing the inclusive 
production of level 2 gauge bosons in UED at the Tevatron 
and the LHC~\cite{KMTev, KongTev, KMAPS, KongPheno}.
We concentrate on the case of $Z_2$ and $\gamma_2$ 
(the level 2 KK partners of the $Z$-boson and the photon, correspondingly),
which give clean dilepton signatures. (In contrast, the discovery of the 
level 2 KK gluon $g_2$ in a dijet channel appears very challenging.)
In the left panel of Fig.~\ref{fig:reach_B2} we show the $5\sigma$ 
discovery reach of the Tevatron and the LHC for 
$\gamma_2$~\cite{KMTev, KongTev, KMAPS, KongPheno}. 
(The reach for $Z_2$ is very similar.) We studied both dielectron and 
dimuon final states, and we plot the required total integrated 
luminosity in ${\rm fb}^{-1}$ as a function of the inverse size of the 
extra dimension $R^{-1}$. We see that already with 10 ${\rm fb}^{-1}$ 
of data (one year of low-luminosity running) the LHC will be able to 
cover all of the cosmologically preferred parameter space of the UED 
model~\cite{UEDdm}. We also see that, just like in supersymmetry, 
there is a significant improvement of the reach once one considers
indirect production of the $\gamma_2$ KK particle, due to 
$n=2$ KK quark decays. Unfortunately, even if the $\gamma_2$ and $Z_2$
are discovered, they may still be misinterpreted as ordinary $Z'$ 
gauge bosons in extended supersymmetric models.
It is therefore necessary to get an independent confirmation
of the discovery of an extra dimension by other means.

\subsection{Spin determinations}
\label{sec:spin}

The second fundamental distinction between UED and supersymmetry is 
reflected in the properties of the individual particles: the
KK partners have identical spin quantum numbers as their SM counterparts, 
while the spins of the superpartners differ by $1/2$ unit. 
However, spin determinations appear to be difficult at the LHC
(or at hadron colliders in general), where the center of mass energy
in each event is unknown. In addition, the momenta of the two dark
matter candidates in the event are also unknown. Recently it has been
suggested that a charge asymmetry in the lepton-jet invariant mass
distributions from a particular cascade, can be used to discriminate
SUSY from the case of pure phase space decays~\cite{Barr:2004ze}
and is an indirect indication of the superparticle spins.
It is therefore natural to ask whether this method can be
extended to the case of SUSY versus UED discrimination.

\begin{figure*}[t]
\begin{center}
%$\begin{array}{c@{\hspace{1cm}}c}
\unitlength=1.0 pt
\SetScale{1.0}
\SetWidth{0.7}      % line    size control
%\unitlength=0.7 pt
%\SetScale{0.7}
%\SetWidth{0.7}      % line    size control
\footnotesize    %  letter  size control
%\qquad
%\allowbreak
%
%  diagram # 2
\begin{picture}(200,100)(0,0)
\SetColor{Red}
\Text(  0.0,79.0)[r]{\red{SUSY:}}
\Text( 20.0,79.0)[l]{\red{$\tilde q$}}
\Text( 90.0,65.0)[r]{\red{$\tilde \chi^0_2$}}
\Text(130.0,45.0)[r]{\red{$\tilde \ell^\mp_L$}}
\Text(175.0,15.0)[l]{\red{$\tilde \chi^0_1$}}
\DashLine(10.0,70.0)(50.0,70.0){3}
\Line(50.0,70.0)(90.0,50.0)
\DashLine(90.0,50.0)(130.0,30.0){3}
\Line(130.0,30.0)(170.0,10.0)
\SetColor{Blue}
\Text(  0.0,55.0)[r]{\blue{UED:}}
\Text( 20.0,55.0)[l]{\blue{$Q_1$}}
\Text( 65.0,45.0)[r]{\blue{$Z_1$}}
\Text(105.0,25.0)[r]{\blue{$\ell^\mp_1$}}
\Text(175.0, 0.0)[l]{\blue{$\gamma_1$}}
\Line(10.0,65.0)(50.0,65.0)
\Photon(50.0,65.0)(90.0,45.0){3}{6}
\Line(90.0,45.0)(130.0,25.0)
\Photon(130.0,25.0)(170.0,5.0){3}{6}
\SetColor{Green}
\Text( 95.0,90.0)[l]{\green{$q$}}
\Text(135.0,70.0)[l]{\green{$\ell^\pm$ (near)}}
\Text(175.0,50.0)[l]{\green{$\ell^\mp$ (far)}}
\Line( 50.0,70.0)( 90.0,90.0)
\Line( 90.0,50.0)(130.0,70.0)
\Line(130.0,30.0)(170.0,50.0)
\end{picture}
% END OF DIAGRAMS
%&
%\includegraphics[width=80mm]{asymmetry.ps}
%\end{array}$
\end{center}
\caption{Twin diagrams in SUSY and UED. The upper (red) line corresponds 
to the cascade decay $\tilde q \to q\tilde\chi^0_2 \to q\ell^\pm\tilde\ell^\mp_L
\to q\ell^+\ell^-\tilde\chi^0_1$ in SUSY. The lower (blue) line corresponds 
to the cascade decay $Q_1 \to q Z_1\to q\ell^\pm\ell^\mp_1
\to q\ell^+\ell^-\gamma_1$ in UED. In either case the observable 
final state is the same: $q\ell^+\ell^-\met$.}
\label{fig:diagrams}
\end{figure*}
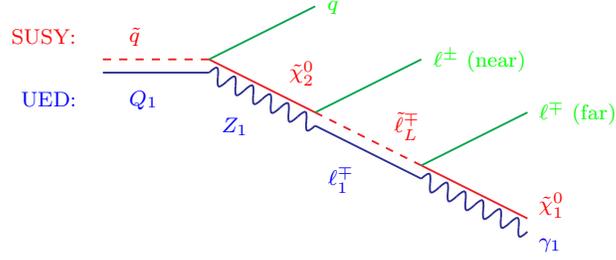

To answer this question, we first choose a study point in UED with
$R^{-1}=500$ GeV. Then we adjust the relevant MSSM parameters until
we get a matching spectrum. Following~\cite{Barr:2004ze}, we concentrate on
the cascade decay $\tilde q \to q\tilde\chi^0_2 \to q\ell^\pm\tilde\ell^\mp_L
\to q\ell^+\ell^-\tilde\chi^0_1$ in SUSY and the analogous decay chain
$Q_1 \to q Z_1\to q\ell^\pm\ell^\mp_1\to q\ell^+\ell^-\gamma_1$ in 
UED~\cite{KMAPS, KongPheno}.
Both of these processes are illustrated in Fig.~\ref{fig:diagrams}.

We then construct the charge asymmetry~\cite{Barr:2004ze}
\begin{equation}
A^{+-} = \frac{\sigma(q\ell^+)-\sigma(q\ell^-)}
              {\sigma(q\ell^+)+\sigma(q\ell^-)}\ ,
\label{asymmetry}
\end{equation}
where $q$ stands for both a quark and an antiquark. The comparison between the
case of UED and SUSY~\cite{KMAPS, KongPheno} 
is shown in the right panel of Fig.~\ref{fig:reach_B2}.
We see that although there is some minor difference in the shape of the 
asymmetry curves, overall the two cases appear to be very difficult to
discriminate unambiguously, especially since the regions near the two ends of the
plot, where the deviation is the largest, also happen to suffer from
poorest statistics. These results have been recently confirmed 
in~\cite{Smillie:2005ar}.

\section{SUSY-UED DISCRIMINATION AT LEPTON COLLIDERS}
\label{sec:clic}

In order to contrast SUSY and UED at lepton colliders, we consider 
an identical final state in each case: $\mu^+\mu^-$ and missing energy. 
This signature may arise either from KK muon production in UED
\begin{equation}
e^+e^- \to \mu^+_1 \mu^-_1 \to \mu^+ \mu^- \gamma_1 \gamma_1\, ,
\label{mu1}
\end{equation}
or from smuon pair production in supersymmetry:
\begin{equation}
e^+e^- \to \tilde \mu^+ \tilde \mu^- \to \mu^+ \mu^- \tilde \chi^0_1 \tilde \chi^0_1\, .
\label{smuon}
\end{equation}
Again, we choose a UED study point with $R^{-1}=500$ GeV, and adjust 
the SUSY parameters until the two spectra match. For definiteness we 
fix the collider center-of-mass energy at $3$ TeV as is the case of CLIC.

\subsection{Muon angular distributions}

In the case of UED, the KK muons are fermions and their angular distribution is given by
\begin{equation}
\left(\frac{d\sigma}{d\cos\theta}\right)_{UED}\ \sim\ 
1+\frac{E^2_{\mu_1}-M^2_{\mu_1}}{E^2_{\mu_1}+M^2_{\mu_1}}\,\cos^2\theta
\ \longrightarrow\ 1+\cos^2\theta. 
\label{ang_ued}
\end{equation}
As the supersymmetric muon partners are scalars, the corresponding angular 
distribution is
\begin{equation}
\left(\frac{d\sigma}{ d\cos\theta}\right)_{SUSY} \sim 1-\cos^2\theta.
\label{ang_susy}
\end{equation}

\begin{figure*}[t]
\centering
\includegraphics[width=80mm]{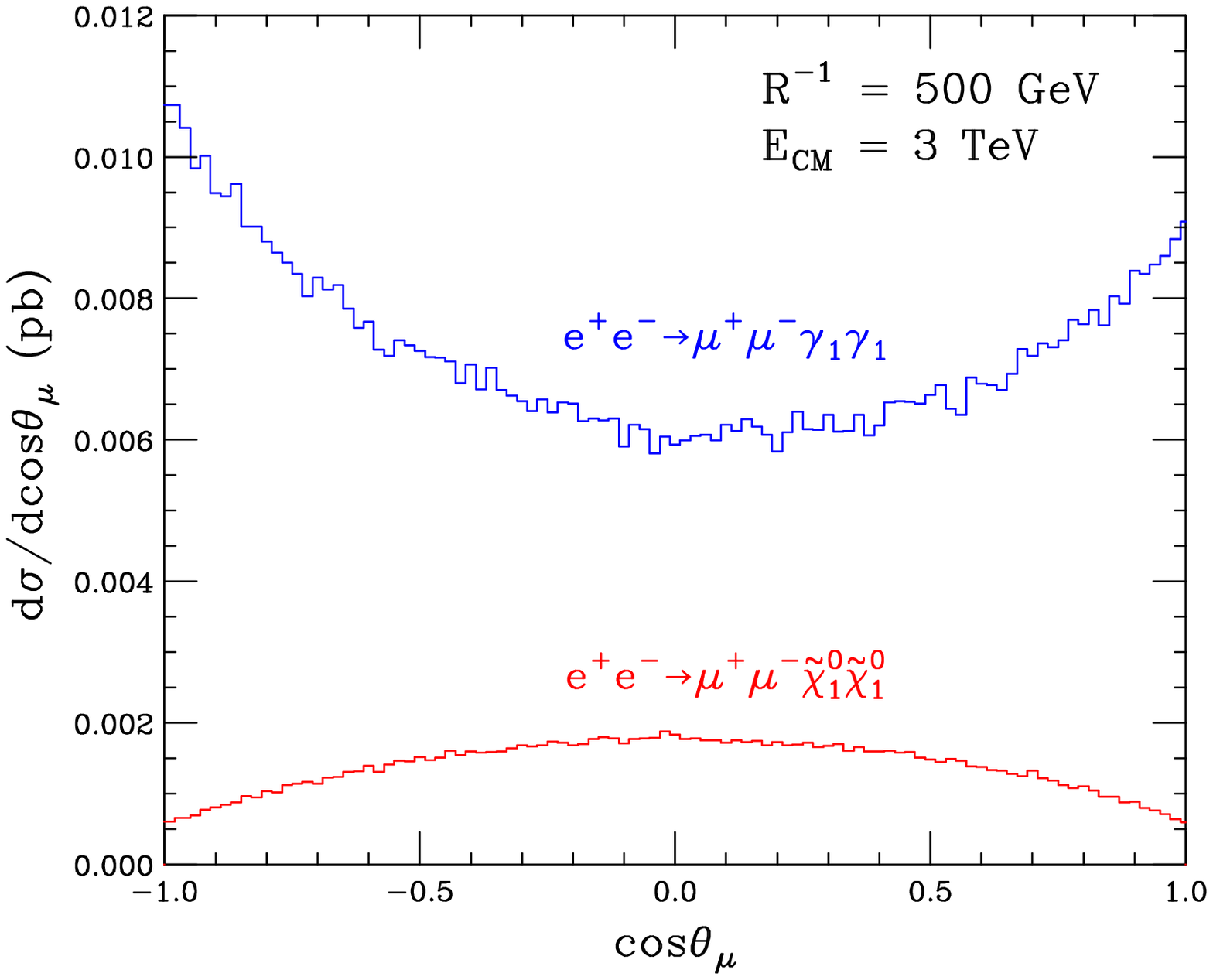}
\hspace{5mm}
\includegraphics[width=70mm]{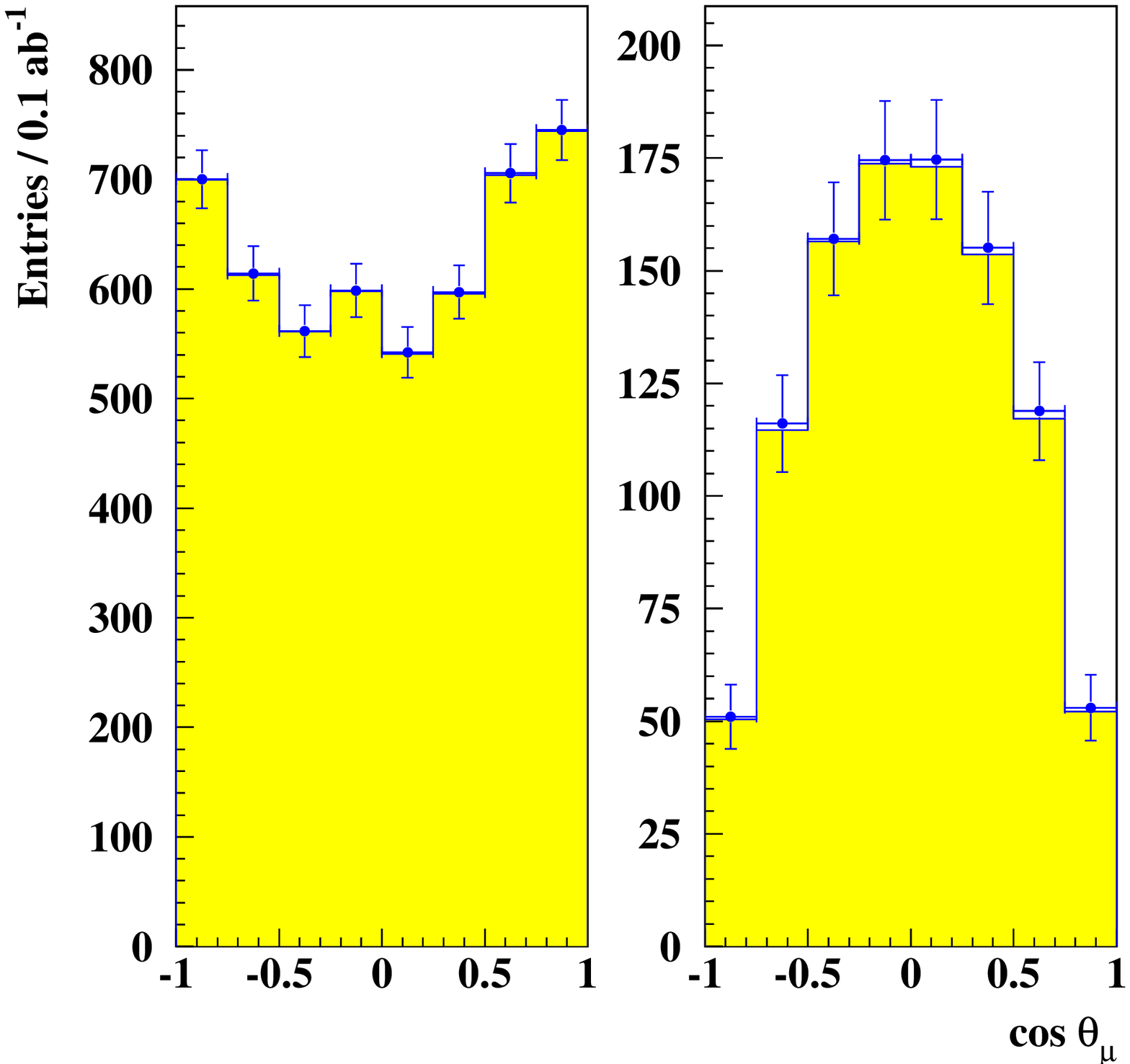}
\caption{Differential cross-section $d\sigma/d\cos\theta_\mu$ 
for UED (blue, top) and supersymmetry (red, bottom)
as a function of the muon scattering angle $\theta_\mu$.
The figure on the left shows the ISR-corrected theoretical prediction.
The two figures on the right in addition include the effects 
of event selection, beamstrahlung and detector resolution and 
acceptance. The left (right) panel is for the case of UED (supersymmetry).
The data points are the combined signal and background events, while the 
yellow-shaded histogram is the signal only.}
\label{fig:ang}
\end{figure*}

Distributions (\ref{ang_ued}) and (\ref{ang_susy}) are sufficiently distinct to 
discern the two cases. However, the polar angles $\theta$ of the original KK-muons 
and smuons are not directly observable and the production polar angles
$\theta_\mu$ of the final state muons are measured instead. But as long as the mass 
differences $M_{\mu_1}-M_{\gamma_1}$ and $M_{\tilde\mu}-M_{\tilde\chi^0_1}$ respectively
remain small, the muon directions are well correlated with those of their parents 
(see Figure~\ref{fig:ang}a).
In Fig.~\ref{fig:ang}b we show the same comparison after detector simulation and 
including the SM background. The angular distributions are well distinguishable 
also when accounting for these effects. It is also clear that the total 
cross-section in each case is very different and provides an alternative discriminator
between the models.

\subsection{Muon energy distributions}

\begin{figure*}[t]
\centering
\includegraphics[width=80mm]{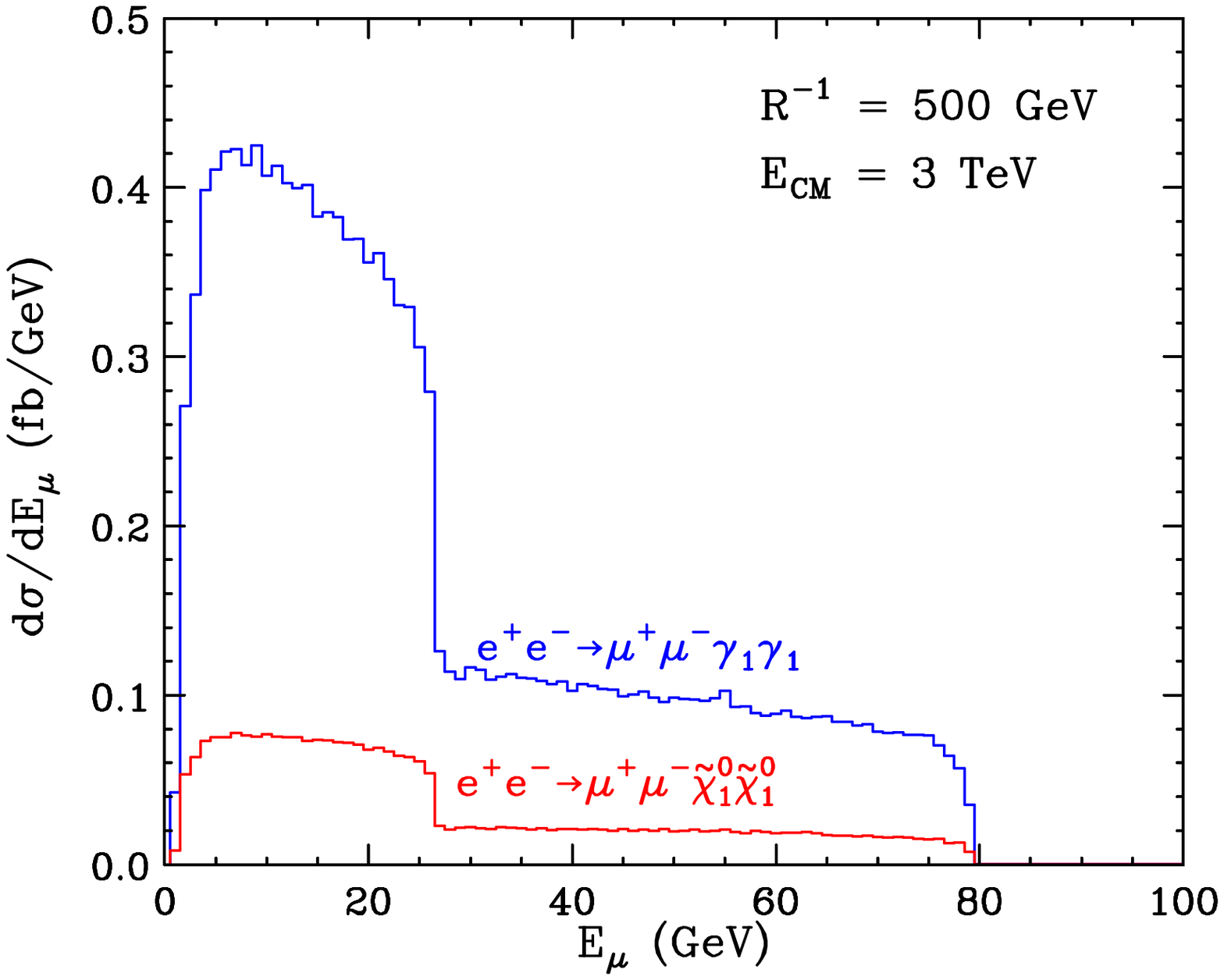}
\hspace{5mm}
\includegraphics[width=70mm]{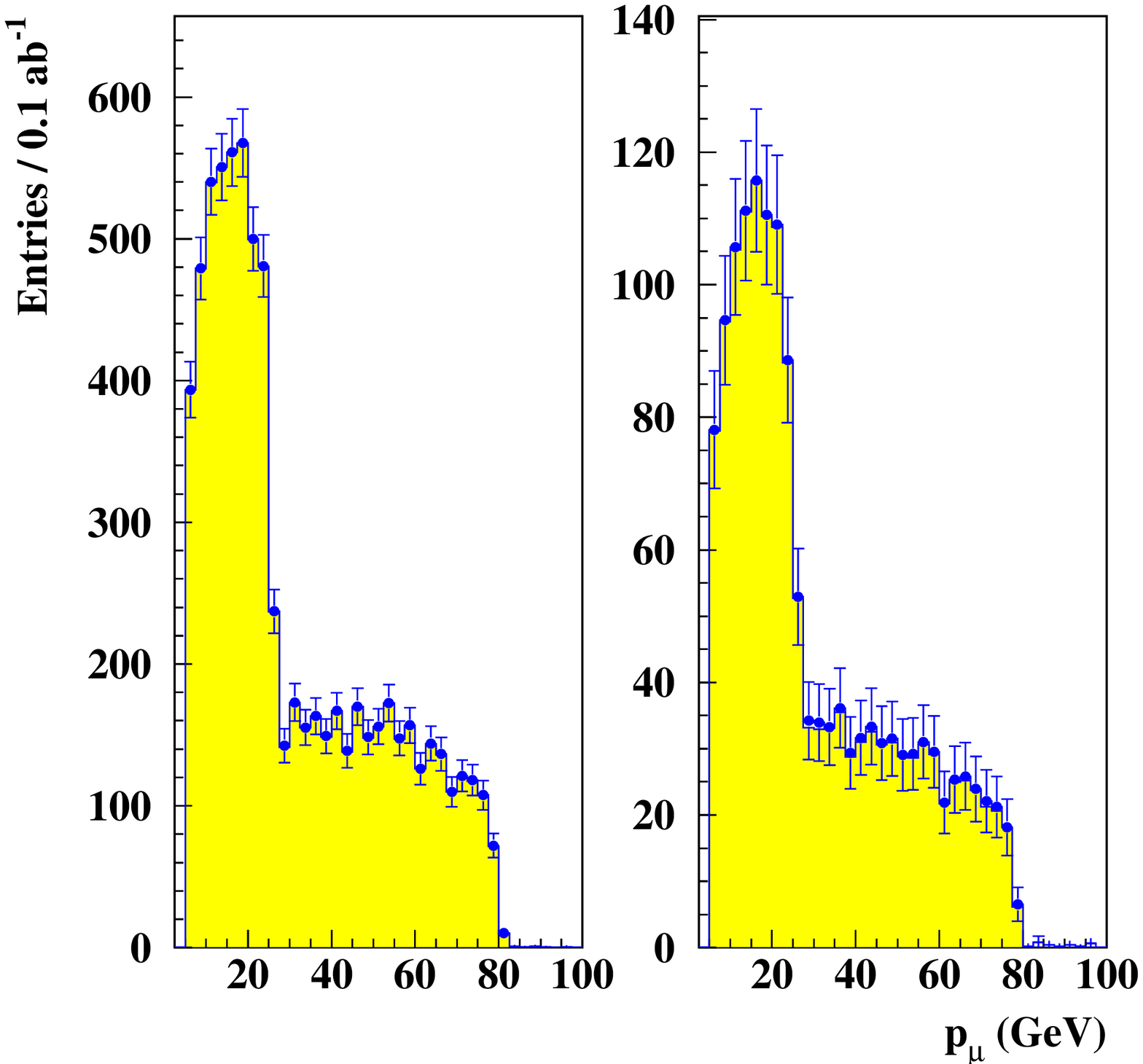}
\caption{The muon energy spectrum resulting from KK muon production (\ref{mu1})
in UED (blue, top curve) and smuon production (\ref{smuon}) in supersymmetry
(red, bottom curve). The UED and SUSY parameters are chosen as in Fig.~\ref{fig:ang}. 
The plot on the left shows the ISR-corrected distribution,
while that on the right includes in addition the effects of event selection, beamstrahlung 
and detector resolution and acceptance. The data points are the combined signal 
and background events, while the yellow-shaded histogram is the signal only.}
\label{fig:emu}
\end{figure*}

The characteristic end-points of the muon energy spectrum are completely determined
by the kinematics of the two-body decay and do not depend on the underlying
framework (SUSY or UED) as long as the spectra are tuned to be identical.
This is illustrated in Fig.~\ref{fig:emu} (we use the same parameters as in
Fig.~\ref{fig:ang}), where we show the ISR-corrected distributions 
for the muon energy spectra at the generator level (left) and after 
detector simulation (right).

The lower, $E_{min}$, and upper, $E_{max}$, endpoints of the muon energy spectrum are 
related to the masses of the particles involved in the decay according to the relation:
\begin{equation}
E_{max/min} = \frac{1}{2} M_{\tilde\mu}
\left(1 - \frac{M^2_{\tilde\chi^0_1}}{M^2_{\tilde\mu}}\right) \gamma (1\pm\beta)\ ,
\end{equation}
where $M_{\tilde\mu}$ and $M_{\tilde \chi^0_1}$ are the smuon and LSP masses and 
$\gamma=1/(1-\beta^2)^{1/2}$ with $\beta = \sqrt{1-M^2_{\tilde\mu}/E^2_{beam}}$ is the 
$\tilde\mu$ boost. In the case of the UED the formula is completely analogous with 
$M_{\mu_1}$ replacing $M_{\tilde \mu}$ and $M_{\gamma_1}$ replacing 
$M_{\tilde \chi^0_1}$.

Due to the splitting between the $\tilde \mu_L$ and $\tilde \mu_R$ masses in MSSM and 
that between the $\mu_1^D$ and $\mu_1^S$ masses in UED, in Fig.~\ref{fig:emu} we see 
the superposition of two box distributions. The left, narrower distribution is due to
$\mu^S_1$ pair production in UED ($\tilde\mu_R$ pair production in supersymmetry).
The underlying, much wider box distribution is due to $\mu^D_1$ pair production in UED 
($\tilde\mu_L$ pair production in supersymmetry). The upper edges are well defined, 
with smearing due to beamstrahlung and, but less importantly, to momentum 
resolution. 
%The lower end of the spectrum has the overlap of the two contributions 
%and with the underlying background. 
%Furthermore, since the splitting between the 
%masses of the $\mu_1^D$, $\mu_1^S$ and that of $\gamma_1$ is small, the lower end 
%of the momentum distribution can be as low as $\cal{O}$(1~GeV) where the lepton 
%identification efficiency is cut-off by the solenoidal field bending the lepton
%before it reaches the electro-magnetic or the hadron 
%calorimeter~\cite{Battaglia:2003wh}.  
Nevertheless, 
there is sufficient information in this distribution to extract the mass of 
the $\gamma_1$ particle, given the values of the $\mu_1^D$ and 
$\mu_1^S$ masses, which in turn can be obtained by a threshold scan.

%In Fig.~\ref{fig:emu}b we show the muon energy distribution after detector simulation.
%A one parameter fit gives an uncertainty on the $\gamma_1$ mass of 
%$\pm$0.19~(stat.)~$\pm$0.21~(syst)~GeV, where the statistical uncertainty is given 
%for 1~ab$^{-1}$ of data and the systematics reflects the effect of the 
%uncertainty on the $\mu_1$ masses. The beamstrahlung introduces an additional 
%systematics, which depends on the control of the details of the luminosity spectrum.

\subsection{Radiative return photon}

With the $e^+e^-$ colliding at a fixed center-of-mass energy above the pair production  
threshold, a significant fraction of the KK muon production will proceed through 
radiative return. Since this is mediated by $s$-channel narrow resonances, a sharp peak 
in the photon energy spectrum appears whenever one of the mediating $s$-channel 
particles is on-shell. In case of supersymmetry, only $Z$ and $\gamma$ particles can 
mediate smuon pair production and neither of them can be close to being on-shell.
On the contrary, an interesting feature of the UED scenario is that $\mu_1$ production 
can be mediated by $Z_2$ and $\gamma_2$ exchange. Since the decay 
$Z_2\to \mu_1\mu_1$ is allowed by phase space, there will be a sharp peak 
in the photon spectrum, due to a radiative return to the $Z_2$.
The photon peak is at
\begin{equation}
E_\gamma = \frac{1}{2}\ E_{CM}\ 
\biggr(1-\frac{M^2_{Z_2}}{ E^2_{CM}} \biggl).
\end{equation}
On the other hand, $M_{\gamma_2}<2M_{\mu_1}$, so that
the decay $\gamma_2\to\mu_1\mu_1$ is closed, 
and there is no radiative return to $\gamma_2$.

\begin{figure*}[t]
\centering
\includegraphics[width=80mm]{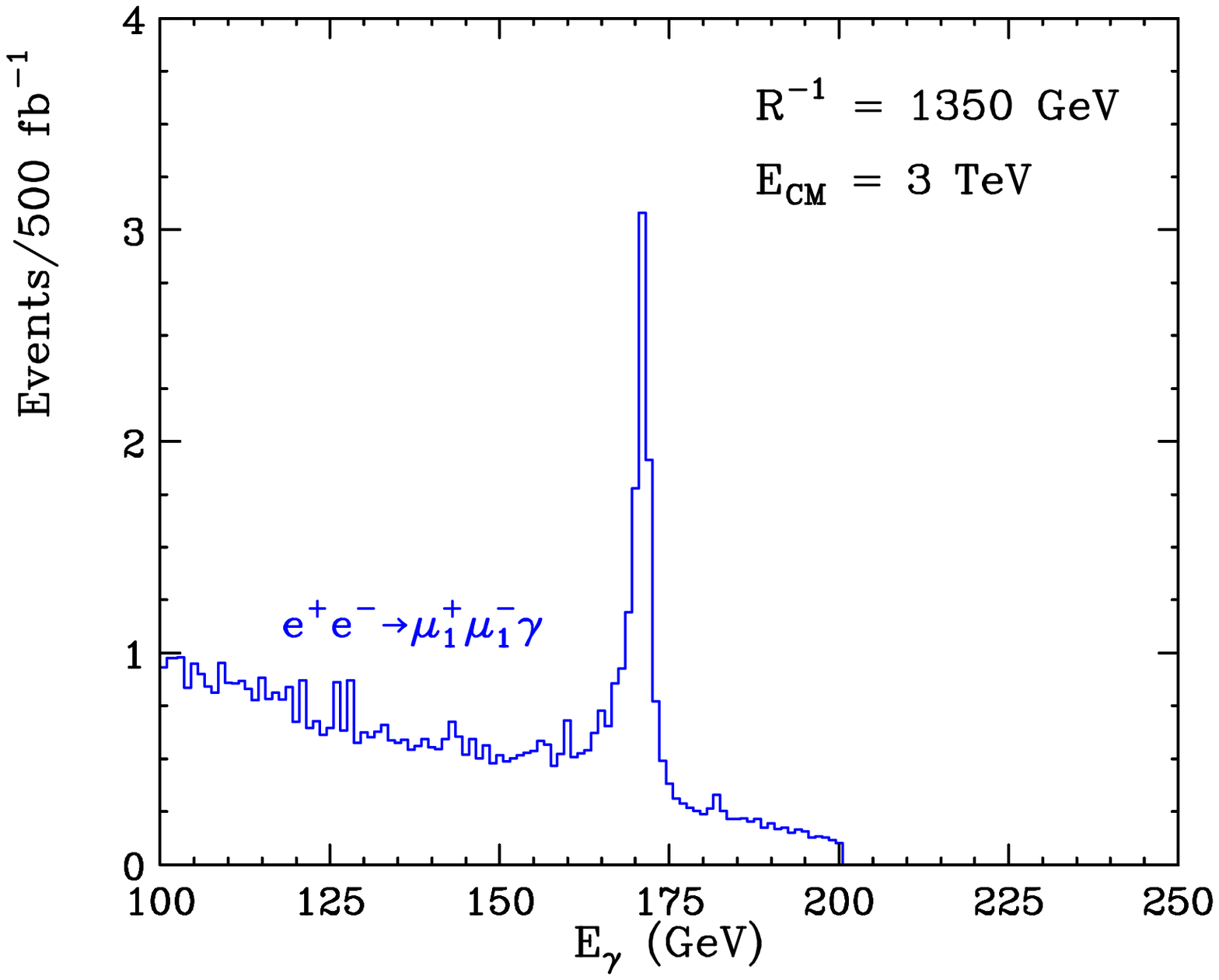}
\hspace{5mm}
\includegraphics[width=70mm]{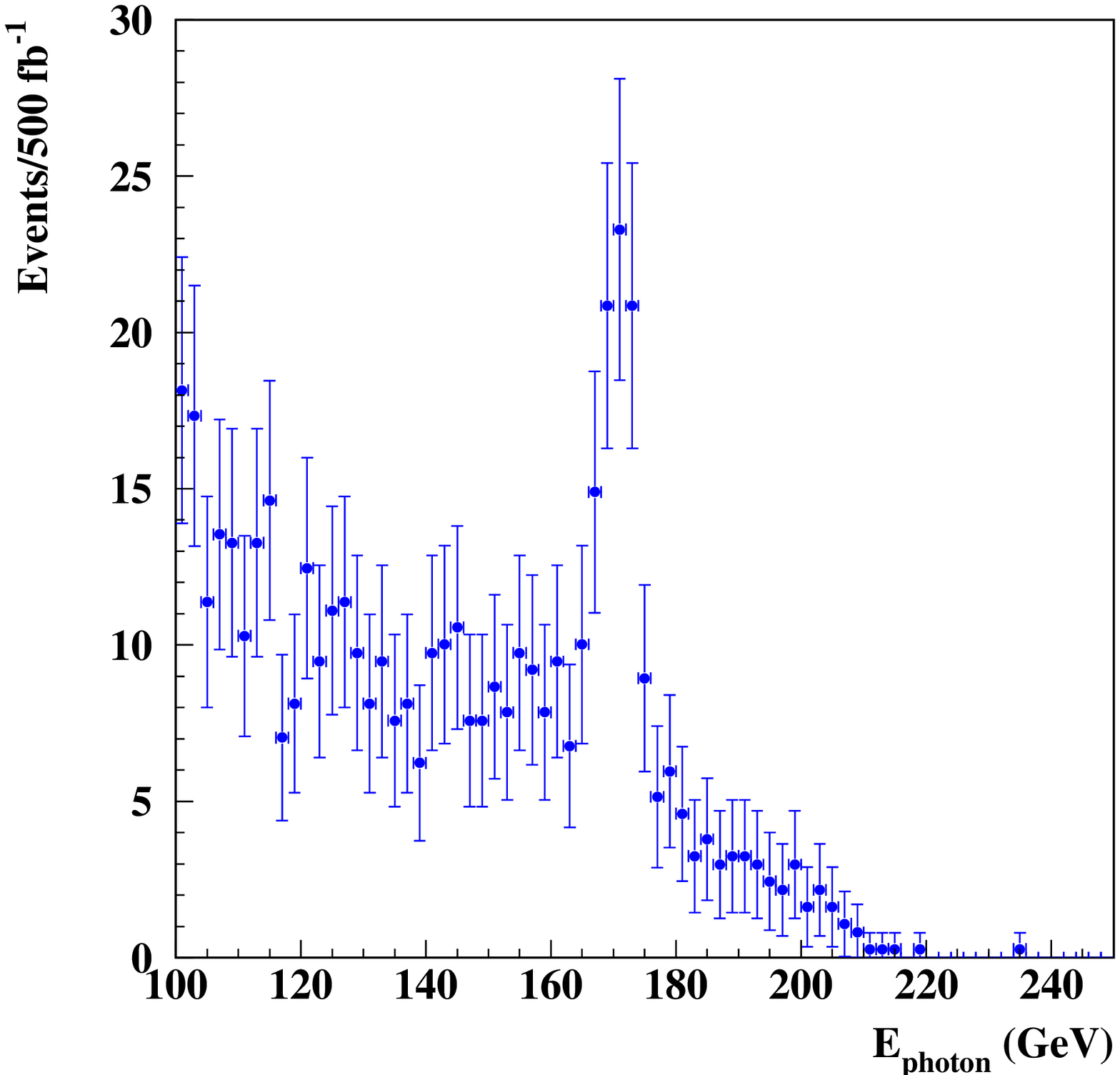}
\caption{Photon energy spectrum in $e^+e^-\to \mu^+_1\mu^-_1\gamma$
for $R^{-1}=1350$ GeV, $\Lambda R=20$ and $E_{CM}=3$ TeV before (left) 
and after (right) detector simulation.
The acceptance cuts are $E_\gamma>10$ GeV and $1<\theta_\gamma<179^\circ$.
The mass of the $Z_2$ resonance is 2825 GeV.}
\label{fig:egamma}
\end{figure*}

The photon energy spectrum in $e^+e^-\to \mu^+_1\mu^-_1\gamma$
for $R^{-1}=1350$ GeV, $\Lambda R=20$ and
$E_{CM}=3$ TeV is shown in Fig.~\ref{fig:egamma}.
On the left we show the ISR-corrected theoretical prediction
from {\tt CompHEP}~\cite{Pukhov:1999gg} while 
the result on the right in addition includes
detector and beam effects. It is clear that the peak cannot be missed.

\begin{acknowledgments}
AD is supported by the US DoE and the 
Michigan Center for Theoretical Physics.
The work of KK and KM is supported in part by 
a US DoE Outstanding Junior Investigator 
award under grant DE-FG02-97ER41209.
\end{acknowledgments}

%\begin{thebibliography}{9}   % Use for  1-9  references

\end{document}